# Quenched narrow-line second- and third-stage laser cooling of $^{40}$Ca

E. Anne Curtis*, Christopher W. Oates and Leo Hollberg

*National Institute of Standards and Technology*

325 Broadway, Boulder, CO 80305

We demonstrate three-dimensional (3-D) quenched narrow-line laser cooling and trapping of $^{40}$Ca. With 5 ms of cooling time we can transfer 28 % of the atoms from a magneto-optic trap based on the strong 423 nm cooling line to a trap based on the narrow 657 nm clock transition (that is quenched by an intercombination line at 552 nm), thereby reducing the atoms' temperature from 2 mK to 10 µK. This reduction in temperature should help reduce the overall systematic frequency uncertainty for our Ca optical frequency standard to < 1 Hz. Additional pulsed, quenched narrow-line third-stage cooling in 1-D yields sub-recoil temperatures as low as 300 nK, and makes possible the observation of high-contrast two-pulse Ramsey spectroscopic lineshapes.

# Introduction

Alkali atoms are routinely laser cooled to temperatures ranging from 1-20 µK (corresponding to several atomic recoils) by the use of optical molasses on strong transitions. In an effort to further reduce atom temperatures, investigators developed new methods to push the cooling below the recoil limit, including the realization of several new laser-cooling strategies: Raman cooling[1,2], velocity-selective coherent population trapping[3], and evaporative cooling in dipole traps[4]. Such second-stage cooling schemes have attained temperatures well below 1 µK and have been used in a variety of experimental demonstrations.

Second-stage cooling is particularly important for alkaline-earth elements since magneto-optic traps (MOTs) involving these atoms typically have temperatures of several millikelvins. These relatively high temperatures are the consequence of Doppler-limited cooling on strong transitions that lack the ground-state substructure required for sub-Doppler cooling mechanisms such as polarization-gradient cooling. There is however real interest in obtaining colder alkaline-earth atoms, since this would benefit many applications including atom interferometry, quantum degenerate gases, and optical frequency standards. In fact, recent frequency measurements of the Ca clock transition at 657 nm were limited (at ~10 Hz) primarily by residual Doppler effects due to the millikelvin temperatures of the laser-cooled atoms.[5,6]

In an effort to reduce the temperature of our Ca atoms, we looked to cooling experiments in Sr, whose narrow intercombination line at 689 nm had been used to build a second-stage MOT and achieved sub-microkelvin temperatures.[7,8] The corresponding transition in Ca (400 Hz linewidth), while excellent for optical frequency standards, cycles photons too slowly to use directly for second-stage cooling of atomic samples starting with a temperature of several millikelvins. However, by incorporating a technique first demonstrated for sideband-cooling with trapped ions[9], we are able to accelerate the cooling process by *quenching* the excited state, that is, connecting the excited state with another laser to a higher-lying level that decays quickly to the ground state. We previously used quenched narrow-line laser cooling (QNLC)[10] in 1-D to reduce the atomic temperature to 4 µK, thus showing the potential of the technique. For the Ca optical frequency standard, however, full 3-D cooling is required to reduce the systematic frequency shifts. Quenched cooling and trapping in 3-D for Ca was first demonstrated at the Physikalisch-Technische Bundesanstalt (PTB), where a quenching transition at 453 nm was used to achieve temperatures of less than 10 µK and transfer efficiencies of ~12 %.[11]

Here we describe 3-D second-stage cooling of Ca, in which we use a different quenching transition to achieve similar temperatures but higher transfer efficiency (~25 %). These microkelvin temperatures should enable the Ca optical frequency standard/clock at 657 nm to achieve a frequency uncertainty of < 1 Hz (fractionally < $3 \times 10^{-15}$). Additionally, we use this ultracold sample as a starting point for third-stage QNLC in 1-D; with a step-wise excitation scheme based on our earlier work[10] we can now produce sub-recoil temperatures as low as 300 nK for 60 % of the ultracold atoms. These sub-microkelvin atomic samples derived from our three-stage cooling scheme are then used for the first demonstration of two-pulse Ramsey fringes with Ca atoms.

## 3-D Second-stage Cooling

*Experiment*

While the step-wise excitation approach we used for 1-D cooling can be extended to 3-D[2], it is more straightforward to implement the method demonstrated in Ref. 11 for 3-D cooling and trapping. This approach uses simultaneous rather than sequential excitation of the three-level system (clock + quenching transitions).  One of the powerful aspects of this technique results from the observation that the three-level system can be reduced to an equivalent two-level description with an effective linewidth approximately equal to the quenching rate.  Thus, the linewidth of the effective Doppler cooling transition can be tuned simply by changing the quenching power.  One can then envision a cooling procedure that starts with a broad transition linewidth (i.e., with its wide Doppler coverage and fast cooling rate) to cool a large range of velocities and then concludes with reduced quenching power for narrow-line cooling and its associated colder temperatures.  For cooling on the clock transition of Ca, however, achieving high quenching rates is difficult since one necessarily must use a weak (intercombination) line to pump atoms from the $^3P_1$ level back to the singlet manifold for quick decay to the ground state.  We have chosen to use the $^3P_1 \rightarrow {}^1S_0$ (4s4p→4s5s) transition at 552 nm (see Figure 1) rather than the $^3P_1 \rightarrow {}^1D_2$ (4s4p→4s4d) transition at 453 nm used in Ref. 11.  The two transitions have similar Einstein A coefficients, but the slower (by a factor of two) decay from the $^1D_2$ state actually accelerates the pumping process, since the pumping rate is proportional to $1/\Gamma$, where $\Gamma$ is the decay rate of the upper state in the three level system.  However, the advantage of using 453 nm pumping is offset by the fact that the dyes used in the green lasers are generally more powerful and longer-lived than those in the blue.  In either case, typical quenching times achieved thus far range from 25 to 50 μs, corresponding to cooling linewidths of < 10 kHz.  These effective cooling linewidths are too narrow to realize the idealized tunable cooling as described above, so we broaden the spectrum of the red laser to extend the capture range of the second-stage cooling light.[7,8,11]

Our cooling route to microkelvin temperatures (see timing diagram, Figure 2) begins with first-stage laser trapping on the strong $^1S_0 \rightarrow {}^1P_1$ transition at 423 nm.  The trapping apparatus has been described elsewhere[12], so we describe only the most relevant details here. A frequency-doubled diode-laser system generates 40 mW of trapping light at 423 nm, which we use to load ~$2\times10^6$ atoms in 25 ms from an atomic beam into a MOT. The resulting temperature of the trapped atoms is 2-3 mK. We follow this with a cycle of second-stage cooling, which uses simultaneous 657 nm cooling and 552 nm quenching light. The 552 nm light comes from a dye laser whose beam, after passing through an optical fiber and liquid-crystal light shutter, supplies ~100 mW of light to the atoms. For long-term stability the frequency of the green light is stabilized relative to a hyperfine component of a nearly coincident $I_2$ transition.

The 657 nm cooling light is generated by an extended-cavity diode laser whose output is amplified by injection-locking two slave lasers. The output from one slave passes through a polarization-maintaining fiber and is used for second-stage cooling and trapping. The other slave supplies light to two separate polarization-maintaining fibers that provide counter-propagating beams for spectroscopy and pulsed cooling (see Refs. 10, 12 for spectroscopic details). Since the red light is used for sub-kilohertz spectroscopic investigations in addition to the cooling described here, it has extremely high frequency stability. This is achieved by locking its frequency tightly to a narrow resonance of a high-finesse Fabry-Perot cavity (linewidth = 9 kHz). The cavity is vibrationally- and acoustically-isolated, and the addition of temperature stabilization of the box surrounding the cavity has led to drift rates of only a few hertz per second. For cooling, the red light is detuned 850 kHz (±50 kHz) below resonance and broadened to cover a 1.5 MHz wide spectrum by applying frequency modulation at 20 kHz to a 72 MHz acousto-optic modulator (AOM). Deviations of 20 % for these modulation parameters did not noticeably affect the cooling results. This modulation creates a comb of optical frequencies close enough together that the atoms remain resonant as they are cooled. The comb width enables a larger portion of the initial velocity distribution of Ca atoms to be cooled and trapped than with a single frequency alone. The trade-off for this broader spectrum is a less precisely tailored spectrum around zero velocity, so we do not expect to attain the temperatures achievable with single-frequency or pulsed cooling.

The red cooling beams (with an intensity of ~20 mW/cm$^2$ per beam and 1/e$^2$ diameter of 3 mm) are retro-reflected in three orthogonal directions and overlap the blue trapping beams. A single green beam (with an intensity of ~ 1 W/cm$^2$ and 1/e$^2$ diameter of 3 mm) has a separate retro-reflected path that lies along two orthogonal axes. The polarization of the red beams matches that of the blue trapping beams, which have the usual $\sigma^+$-$\sigma^-$ configuration for a magneto-optic trap. The green beams, however, have the opposite circular polarization in order to increase the likelihood of absorption of green and red photons from the same direction, thus increasing the cooling force. This increase results from considerations of angular momentum; since the $^1S_0$ upper state of the quenching transition has only an m=0 level, quenching of the +1 and -1 levels of the $^3P_1$ state necessarily requires absorption of $\sigma^-$ and $\sigma^+$ light, respectively. To add a spatial (i.e. trapping) component to the narrow-line cooling process we introduce a small magnetic quadrupole field (~30 µT/cm or ~300 mG/cm) as has been used in similar alkaline-earth MOTs.[7,8,11] The cooling results were not sensitive to the size of the gradient over the range of 10-50 µT/cm.

The second-stage cooling is followed by a pulse of quenching light (~100 µs) that is used to transfer those atoms remaining in the excited state to the ground state for maximum detection efficiency. We then use a red spectroscopic pulse whose frequency is swept to measure the velocity distribution of the atoms along one direction (one point per measurement cycle) via the first-order Doppler shift. In order to prevent broadening the duration of this pulse is chosen to yield a Fourier-transform width much less than that of the velocity distribution. To take advantage of the presence of the quenching light, we have modified the normalized shelving detection scheme (based on near-resonant blue pulses) used in our previous experiments to achieve a high signal-to-noise ratio.[12] Now, instead of using a blue normalization pulse before the spectroscopy, we perform the normalization at the end of the spectroscopic cycle. The detection process begins with a blue probe pulse that occurs after the red velocity probe pulse in order to measure those atoms left in the ground state (i.e., those not in the velocity class excited by the red probe pulse). Atomic fluorescence excited by the standing-wave blue probe (intensity ~15 mW/cm$^2$, with a duration of 30 μs) is collected by a lens inside the vacuum system and directed to a photomultiplier tube. The excited atoms are then pumped back to the ground state by a quenching pulse (552 nm, 150 μs duration), and the ground-state population (now equal to the total population of trapped atoms) is measured by a second blue normalization pulse. Dividing the integrated fluorescence of the first pulse by that of the second yields a normalized excitation signal. By keeping the time between intensity-stabilized blue probe pulses short (and thus less sensitive to residual fluctuations in intensity), and by normalizing the excitation signal relative to the number of atoms captured in the red/green trap (rather than to the number in the blue trap), we are able to maintain the signal-to-noise ratio attained in the spectroscopy based on millikelvin atoms. A similar scheme was recently used at PTB, using a different version of shelving detection with quenching.[13]

As shown in Figure 3, the temperature of the atom cloud drops rapidly in the first few milliseconds of red/green cooling and then levels off below 9 μK for longer cooling times. The number of atoms transferred from the blue trap to the red/green trap also decreases with trap time (see Figure 4), as those atoms whose velocity is not within the capture range of the trap (~40 cm/s) rapidly escape the observation region. Similar results were seen in the second-stage MOT used at PTB, but with smaller atom-transfer efficiency, which we attribute to less quenching laser power. Figure 5 shows a typical velocity distribution after 15 ms of red/green trapping. The symmetry of the measured final velocity distributions is particularly sensitive to the alignment of the green cooling beam, since an asymmetry of green absorptions leads to an asymmetric force on the atoms. Optimization of the overlap of the retro-reflected beams for both the trapping and quenching light increases the atom number and improves the overall symmetry of the velocity distributions. The temperature limit of 5-10 μK observed for these traps is a result of the recoils associated with the quenching process (recall that each absorption of a green photon is followed by emission of photons at 1.03 μm and 423 nm) as well as the shape of the cooling light spectrum around zero velocity.

## *Improved Optical Frequency Standard*

Our particular motivation for second-stage cooling of Ca is to improve the performance of our optical atomic frequency standard. For this application, an optimized version of this second-stage cooling scheme would ideally have the shortest trapping time (to minimize overall cycle time) and the largest number of atoms (to maximize the signal-to-noise ratio), as well as the coldest temperatures (to attain the highest accuracy). As a compromise, we find that working with a red/green trapping time of 5 ms allows the atom cloud to come into equilibrium (good fit to a Gaussian) at a temperature of 10 μK, with a transfer efficiency of 28 % ($5.6 \times 10^5$ of the originally trapped $2.0 \times 10^6$ atoms.) With such a narrow distribution, nearly all of these atoms contribute to the Bordé-Ramsey interferometry we use for our clock measurements, and the resulting lineshapes are virtually Fourier-transform-limited. In Figure 6 we show a single 100-second scan taken at a resolution of 11.55 kHz using ultracold atoms; the contrast in this Bordé-Ramsey fringe pattern is nearly 40 %. We note that the asymmetry in the fringe envelope is real – computer simulations based on the formalism developed by Bordé *et al.*[14] show the same asymmetric envelope and indicate that it results from atomic recoil shifts. For applications to frequency metrology we actually work at considerably higher resolutions (usually ~1 kHz or less), in order to have a more sharply defined line center. The inset to Figure 6 shows the central portion of a fringe pattern with 1.45 kHz resolution.

Let us now consider the improvements we expect to see in our optical standard based on the parameters we can achieve for our ultracold atomic sample. The most dramatic difference should occur in the overall uncertainty with which we can realize the unperturbed atomic frequency, since we were previously limited by velocity-related frequency shifts. The two main systematic effects relating to velocity result from (1) a non-zero transverse drift velocity of the atomic sample combined with a residual mismatch in the overlap of the Bordé-Ramsey spectroscopic beams, and (2) moving atoms sampling the phase of probe beams with a finite radius of curvature. Second-stage cooling reduces the $v_{rms}$ by a factor of 15, thus reducing the shifts due to these effects to < 1 Hz, a regime where other systematic shifts become significant and will require a more detailed evaluation.[13] Nonetheless, for a Ca standard based on ultracold atoms, a frequency uncertainty of < 0.4 Hz (corresponding to a fractional uncertainty of < $10^{-15}$) certainly seems feasible.

A second critical parameter used to characterize clock performance is stability, which is effectively a measure of the frequency uncertainty that is achieved for a given averaging time. Due to their high line Q's, optical frequency standards have demonstrated superior short-term stability to that of their microwave counterparts.[15] In that work, a comparison between our Ca optical clock (based on millikelvin atoms) and a Hg$^+$ optical clock (based on a single Hg ion) yielded a fractional frequency instability of 7×10$^{-15}$ at 1s, which represents the most stable short-term performance demonstrated with atomic frequency standards. In moving to a clock based on ultracold atoms, we see some trade-offs in terms of short-term stability. The longer loading and preparation time of the atoms leads to a six-fold increase in the measurement-cycle duration, which will effectively increase the averaging period, degrading the short-term stability. Moreover, the number of trapped atoms is reduced by nearly a factor of 10, which also reduces the signal-to-noise ratio. Fortunately, this reduction is mostly offset by the fact that with ultracold atoms basically all the atoms contribute to the fringe structure, rather than the ~30 % we had with the millikelvin atoms. Accounting for the 2-3 times higher contrast achieved with ultracold atoms at sub-kilohertz resolutions, we expect to see little net degradation in short-term frequency stability. This is supported by our initial measurements of the fractional frequency instability of the system based on ultracold atoms, which show an Allan deviation of < 2×10$^{-14}$ at 1s. The dominant residual noise appears to be frequency noise on the laser, which we think results from fluctuations in the length of the optical reference cavity, perhaps aggravated by the Dick Effect (for microkelvin atoms our spectroscopic period is only a few percent of the whole measurement cycle).[16] The amplitude noise on the detection sets a limit of < 10$^{-14}$, so we are confident that with improved cavity isolation and an apparatus better suited for second-stage cooling (with more optical access for capturing more atoms), we should be able to attain a short-term instability of close to 1×10$^{-15}$ at 1s. We note that even this level of instability is still an order of magnitude worse than estimates of what could be achieved with this Ca transition, so there is still plenty of room for improvement.[13,17]

## Additional 1-D Cooling

*Experiment*

With the capability of starting experiments with trapped atoms at 10 µK rather than at 2 mK, we decided to revisit pulsed QNLC in 1-D.[10] Our previous results had been hindered by two factors associated with the initial millikelvin temperature: the transverse velocity of the atoms sets an upper limit of ~2.5 ms for the cooling time, while the width of the velocity distribution required many cooling cycles (i.e., long cooling time) for a majority of the atoms to reach near-zero velocity. With the atoms now starting much closer to zero velocity in 3-D, we could hope to reach sub-recoil temperatures in 1-D for a non-trivial fraction of the atomic sample.

The advantage of using π-pulse excitation on the narrow red transition, followed by a quenching pulse in the green, is that we have greater control over the frequency spectrum of the cooling light. Since the cooling pulses are much shorter than the lifetime of the excited state, the effective excitation spectrum of the light is basically the Fourier transform of the time dependence of the cooling pulse. We use square pulses in the time domain whose frequency is detuned to the red such that the first zero of the resulting $(\sin x/x)^2$ excitation spectrum lies at zero velocity.[1,2,10] This produces an effective dark state around zero velocity in which atoms can accumulate. Since the length of the cooling pulses in time is inversely related to their span in frequency (velocity) space, longer pulses produce a narrower hole around zero velocity and thus colder velocity distributions. The trade-off, however, is that these longer pulses drive a smaller range of velocities towards zero, so it is advantageous to start with a fairly narrow velocity distribution.

A typical three-stage cooling cycle is shown in Figure 7. We start with our usual blue cooling followed by red/green cooling and add a third stage that consists of a series of cooling pulses that drive the atomic velocities toward the near zero-excitation region around zero velocity. After each pair of sequential red pulses (one from each direction and with an intensity chosen to yield a π-pulse area), a quenching pulse follows that sends the atoms around to the ground state for another cooling cycle. We alternate the directional order of the red pulses from cycle to cycle to enhance the symmetry of the cooling process. In order to use the m=0→m=0 magnetic-field-insensitive transition for this third stage cooling, we use linearly-polarized red light along with a magnetic bias field of 200 µT along the same direction. The red cooling pulses are derived from the same laser and optical fibers that generate the spectroscopic pulses. To tailor the pulses, we use switches to send different rf signals to the AOMs controlling the intensity and frequency of the red light. Note that we use the same quenching polarization and spatial configuration as in the second stage, which in our present beam geometry means that green light propagates both perpendicular and parallel to the pulsed red light. As a result of the quenching process, the atoms experience several recoils as they are pumped back to the ground state. Those atoms whose resultant velocity falls near zero will likely avoid re-excitation, while the remaining atoms will undergo another cooling cycle. This process repeats until the atoms have all fallen into the near-zero velocity "dark" state, or by virtue of an unfavorable recoil sequence have landed outside the velocity range covered by the excitation pulses. As before, we use high-resolution 657 nm velocity probe pulses to measure the final width of the velocity distribution along the direction of the 1-D cooling pulses.

In Figure 8 we show the 1-D velocity distribution that results from a series of eight cycles of the third-stage cooling using pulses with 10 µs duration. Each cycle contains a quenched red sequence followed by a quenched red sequence with the pulses in opposite order – see Figure 7. Note that we find that for cooling with pulses of duration < 15 µs, the velocity distributions fit to Lorentzian lineshapes rather than to Gaussians, as was the case for 3-D cooling, so the 1-D results are quoted using the half width, half maximum (HWHM) of the Lorentzian instead of an rms width.[18] For a sequence of eight cycles, the resulting $V_{HWHM}$ of the distribution is 1.3 cm/s (corresponding to a temperature of 840 nK), approximately equal to the recoil limit (for a single 657 nm photon). Interestingly, a majority of the cooling occurs during the first two cycles, during which minimal heating of the transverse directions should occur.

As we increase the length of the pulses beyond 15 μs in order to reach lower temperatures, additional structure begins to appear in the baseline of the velocity distributions as more atoms can access $(\sin x/x)^2$ zeroes that are not centered at zero velocity. These additional "dark" states limit the transfer efficiency into the coldest central peak (similar structure was seen in our original QNLC experiments). To reduce this effect, we implement sequences using pulses of differing lengths (each with its associated detuning). In one case we make the first pulse short so that it transfers a large number of atoms towards zero velocity, and the second pulse long so that it produces a narrow velocity distribution. A sequence consisting of 5 cycles of 15 μs pulses followed by 8 cycles of 25 μs pulses leads to a central feature with a temperature of ~520 nK (see Figure 9). One can see here the atoms accumulating in small peaks around the other zeroes of the $(\sin x/x)^2$ function. We find that, under these conditions, if we allow the atoms to decay directly from the excited state in order to avoid the recoils associated with quenching, fewer atoms escape the capture region. This has however the undesirable property of increasing the cooling time by a factor of four. In an effort to achieve colder temperatures while maintaining good transfer efficiency, we instead employ a sequence that alternates 12 μs pulses and 50 μs pulses for 5 cooling cycles. This combination produces a velocity distribution whose central narrow feature has a temperature of only 300 nK (one fourth of the single photon recoil temperature), and transfers ~60 % of the atoms to the central peak.

*Spectroscopy*

In preliminary experiments the third-stage 1-D cooling did not lead to a marked improvement in the fringe contrast for four-pulse Bordé-Ramsey spectroscopy, since we were already close to the theoretical limit of 50 % contrast. However, this extra cooling stage does reduce the velocity of the atoms to the point where the atoms move only a small fraction of an optical wavelength for times as long as 20 µs. In this case we can obtain Ramsey fringe structure using only two pulses, as in the case of microwave Ramsey spectroscopy. Figure 10 shows two-pulse fringes for a Ramsey time (time between red pulses) of 5 µs. For longer times the fringe contrast degrades as the atoms move a greater fraction of a wavelength, thus adding a spatial component to the interference process that effectively washes out the fringes. Nonetheless, with sub-recoil temperatures, we can observe fringes for Ramsey times beyond 20 µs. Not surprisingly, the fringe contrast is very sensitive to velocity; computer simulations of this process yield velocity widths in very good agreement with our direct measurements of the velocity distributions. Due to their first-order Doppler sensitivity and reduced contrast at high resolution, these two-pulse fringes will not supercede the four-pulse variety for our immediate frequency-standard applications, but they do offer a nice demonstration of the increased coherence of the atomic sample. While this is the first demonstration of two-pulse fringes in Ca, we emphasize that two-pulse fringes have previously been seen in the analogous transition in Sr.[18]

## Conclusions

We demonstrate 3-D second-stage laser cooling and trapping of Ca with fairly high transfer efficiency. This greatly reduces the dominant systematic frequency shifts for the Ca optical frequency standard, but also serves as a good starting place for further experiments. We have shown one example in which we have used pulsed cooling in one dimension to reduce the temperature below the recoil limit. Other possibilities include ramping up the magnetic field gradient to increase the density in preparation for further cooling towards Bose-Einstein condensation.[19] As has been demonstrated with Sr[20], these ultracold traps can serve as a good starting place for loading far-detuned optical lattices. In fact, a promising approach for an optical frequency standard would be to load ultracold atoms into a 1-D optical lattice for sideband cooling on the clock transition and spectroscopic interrogation. Following the concept of Katori *et al.*, we can choose the wavelength for the far-detuned dipole traps such that the Stark shift for both the ground and excited states would be equal.[21] One could thus perform the spectroscopy on the clock transition with the lattice fields on, mimicking the environment usually associated with trapped ions. Under these conditions, one could imagine having virtually a million non-interacting clocks, cooled to the Lamb-Dicke regime, with very small systematic uncertainty and frequency instability.

*Also with the University of Colorado, Boulder, CO, 80309

## References:

1. M. Kasevich and S. Chu, "Laser cooling below a photon recoil with three-level atoms," Phys. Rev. Lett. **69**, 1741-1744 (1992).

2. J. Reichel, F. Bardou, M. Ben Dahan, E. Peik, S. Rand, C. Salomon, and C. Cohen-Tannoudji, "Raman Cooling of Cesium below 3 nK: New Approach Inspired by Lévy Flight Statistics," Phys. Rev. Lett. **75**, 4575-4578 (1995).

3. A. Aspect, E. Arimondo, R. Kaiser, N. Vansteenkiste, C. Cohen-Tannoudji, "Laser Cooling below the One-Photon Recoil Energy by Velocity-Selective Coherent Population Trapping," Phys. Rev. Lett. **61**, 826-829 (1988), and J. Lawall, F. Bardou, B.Saubamea, K. Shimizu, M.

Leduc, A. Aspect, C. Cohen-Tannoudji, "Two-Dimensional Subrecoil Laser Cooling," Phys. Rev. Lett. **73**, 1915-1918 (1994).

4. C. S. Adams, H. J. Lee, N. Davidson, M. Kasevich, and S. Chu, "Evaporative Cooling in a Crossed Dipole Trap," Phys. Rev. Lett. **74**, 3577-3580 (1995).

5. T. Udem, S. A. Diddams, K. R. Vogel, C. W. Oates, E. A. Curtis, W. D. Lee, W. M. Itano, R. E. Drullinger, J. C. Bergquist, and L. Hollberg, "Absolute Frequency Measurements of the $Hg^+$ and Ca Optical Clock Transitions with a Femtosecond Laser," Phys. Rev. Lett. **86**, 4996-4999 (2000).

6. G. Wilpers, *Ein optisches Frequenznormal mit kalten und ultakalten Atomen*, Ph.D. Thesis, University of Hannover, 85 (2002).

7. H. Katori, T. Ido, Y. Isoya, M. Kuwata-Gonokami, "Magneto-Optical Trapping and Cooling of Strontium Atoms down to the Photon Recoil Temperature," Phys. Rev. Lett. **82**, 1116-1119 (1999).

8. K. R. Vogel, T. P. Dineen, A. Gallagher, and J. L. Hall, "Narrow-line Doppler cooling of strontium to the recoil limit," IEEE Trans. Instr. and Meas. **48**, 618-621 (1999).

9. F. Diedrich, J. C. Bergquist, Wayne M. Itano, and D. J. Wineland, "Laser Cooling to the Zero-Point Energy of Motion," Phys. Rev. Lett. **62**, 403-406 (1989).

10. E. A. Curtis, C. W. Oates, and L. Hollberg, "Quenched narrow-line laser cooling of $^{40}$Ca to near the photon recoil limit," Phys. Rev. A **64**, 031403(R) (2001).

11. T. Binnewies, G. Wilpers, U. Sterr, F. Riehle, J. Helmcke, T. E. Mehlstäubler, E. M. Rasel, and W. Ertmer, "Doppler Cooling and Trapping on Forbidden Transitions," Phys. Rev. Lett. **87**, 123002 (2001).

12. C.W. Oates, F. Bondu, R. Fox, and L. Hollberg, "A diode-laser optical frequency standard based on laser-cooled Ca atoms: Sub-kilohertz spectroscopy by optical shelving detection," Eur. J. of Phys. D **7**, 449-460 (1999).

13. G. Wilpers, T. Binnewies, C. Degenhardt, U. Sterr, J. Helmcke, and F. Riehle, "An Optical Clock with Ultracold Neutral Atoms," arXiv:physics/0205049, 16 May 2002.

14. Ch. J. Bordé, Ch. Salomon, S. Avrillier, A. Van Lerberghe, Ch. Bréant, D. Bassi, G. Scoles, "Optical Ramsey fringes with traveling waves," Phys. Rev. A **30**, 1836-1848 (1984).

15. S. A. Diddams, T. Udem, K. R. Vogel, C.W. Oates, E. A. Curtis, W. D. Lee, W. M. Itano, R. E. Drullinger, J. C. Bergquist, and L. Hollberg, "An Optical Clock Based on a Single Trapped 199Hg+ Ion," Science **293**, 825-828 (2001).

16. G. J. Dick, "Local oscillator induced instabilities in trapped ion frequency standards," in *Proc. Precise Time and Time Intervals,* (Redondo Beach, CA, 1987), p. 133-147. See also the Special Issue on the Dick Effect, IEEE Trans. on Ultrason., Ferroelect., Freq. Contr. **45**, 876-905 (1998).

17. L. Hollberg, C. W. Oates, E. A. Curtis, E. N. Ivanov, S. A. Diddams, Th. Udem, H. G. Robinson, J. C. Bergquist, W. M. Itano, R. E. Drullinger, and D. J. Wineland, "Optical frequency standards and measurements," IEEE J. Quantum Electronics **37**, 1502-1513 (2001).

18. Lorentzian lineshapes were also observed for ultracold Sr atoms – as reported by K. R. Vogel, *Laser cooling on a narrow atomic transition and measurement of the two-body cold collision loss rate in a strontium magneto-optical trap*, Ph.D. Thesis, University of Colorado, Boulder, CO, 213 (1999).

19. M. H. Anderson, J. R. Ensher, M. R. Matthews, C. E. Wieman, E. A. Cornell, "Observation of Bose-Einstein Condensation in a Dilute Atomic Vapor," Science **269**, 198-201 (1995).

20. Hidetoshi Katori, Tetsuya Ido and Makoto Kuwata-Gonokami, "Optimal Design of Dipole Potentials for Efficient Loading of Sr Atoms," J. Phys. Soc. Jpn. **68**, 2479-2482 (1999).

21. H. Katori, "Spectroscopy of Strontium Atoms in the Lamb-Dicke Confinement," in *Proc. of the Sixth Symposium of Frequency Standards and Metrology,* Patrick Gill, ed. (St. Andrew's, Scotland, 2001), p. 323-330.

## Figures and Figure Captions

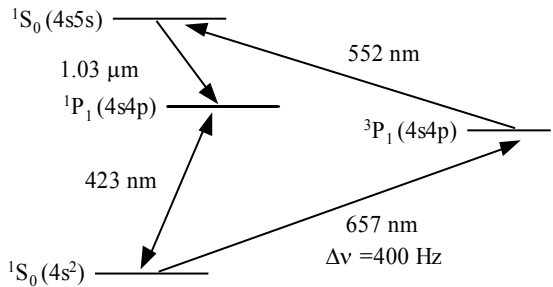

Figure 1: Energy-level diagram for $^{40}$Ca showing relevant levels for cooling and quenching.

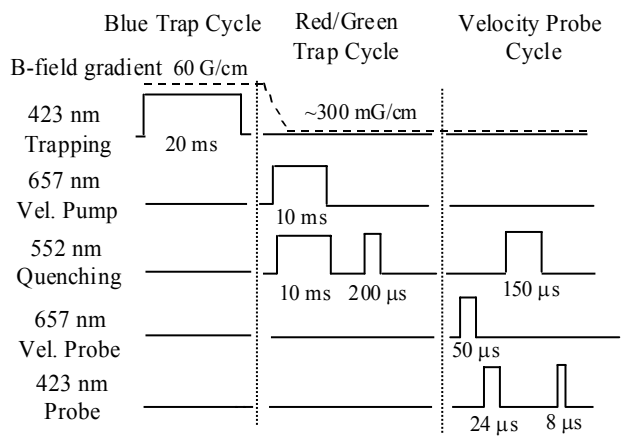

Figure 2: Timing diagram for second-stage cooling. See text for details of the modified shelving detection scheme.

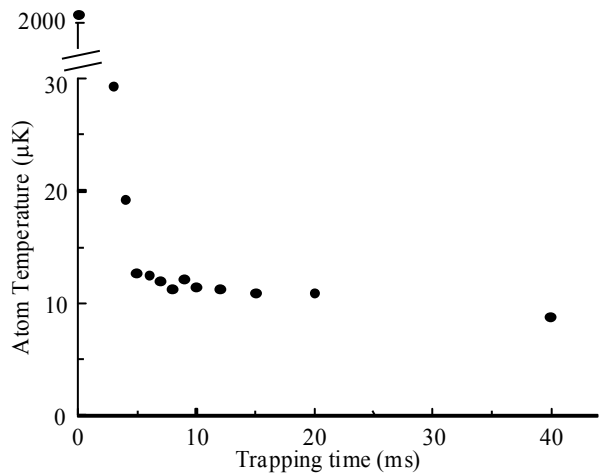

Figure 3: Temperature of the atom cloud (corresponding to the rms velocity derived from the fit of a Gaussian lineshape to the measured velocity distribution) vs. second-stage cooling time. Measurements of the velocity distribution in the other two dimensions yield similar results. The initial trap temperature with no second-stage cooling is ~2000 μK.

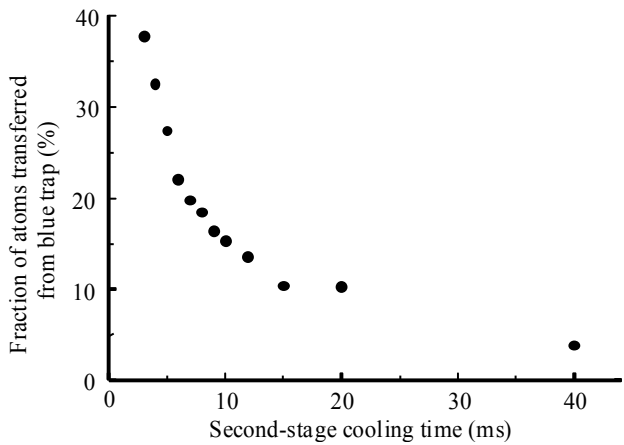

Figure 4: Fraction of atoms transferred from the blue trap to the red/green trap vs. second-stage cooling time.

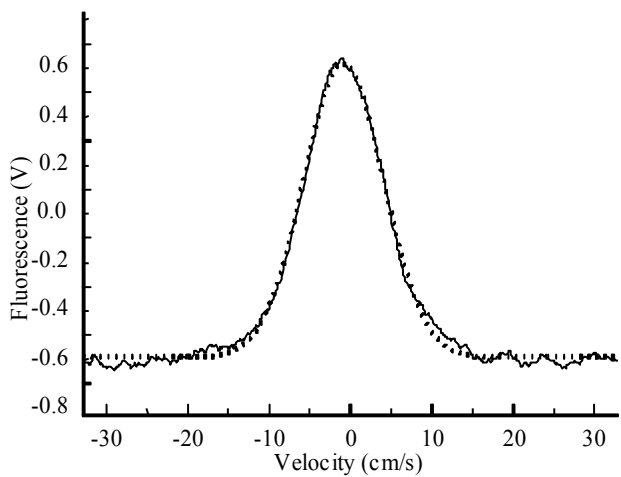

Figure 5: Atomic velocity distribution after 15 ms of red/green cooling and trapping. A fit to a Gaussian gives an rms velocity of 4.9 cm/s, which corresponds to a temperature of 11.5 μK.

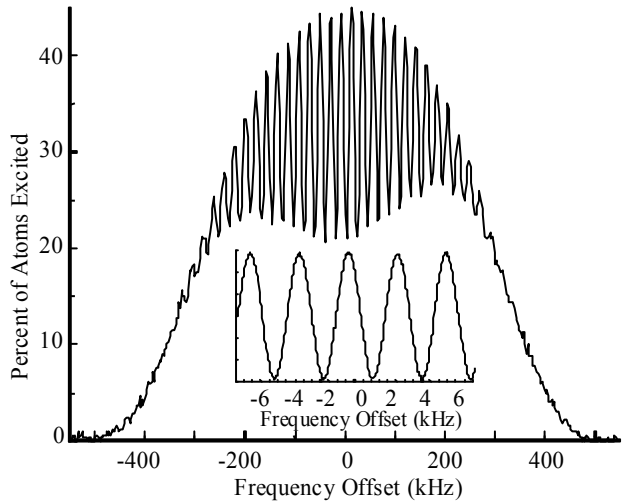

Figure 6: Bordé-Ramsey fringes based on the m=0→m=0 levels of the 657 nm clock transition with a resolution of 11.55 kHz taken after 4 ms second-stage cooling. A single 100-second frequency sweep shows high contrast, Fourier-transform-limited fringes. The asymmetric fringe envelope is a result of atomic recoil. Inset: High resolution 1.45 kHz Bordé-Ramsey fringes taken under the same cooling conditions (< 30 s averaging time.)

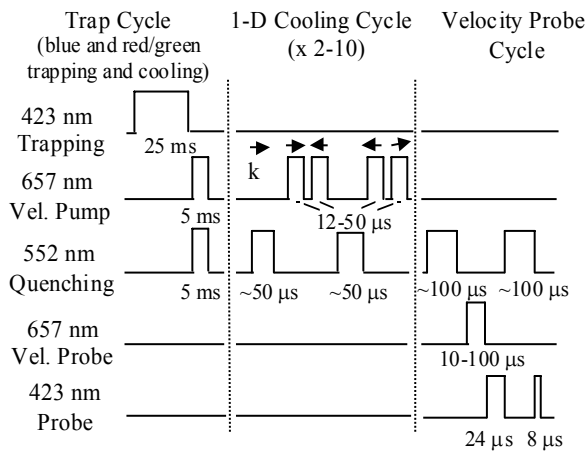

Figure 7: Timing diagram for three-stage cooling showing the additional 1-D narrow-line quenched cooling.

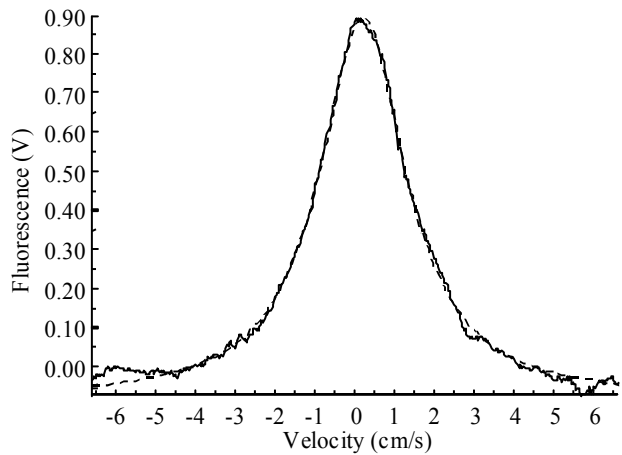

Figure 8: Velocity distribution after 6 ms second-stage cooling followed by 8 cycles of 1-D single-frequency 15 μs red pulses and green quenching. A fit to a Lorentzian yields velocity (HWHM) of 1.31 cm/s (sub-recoil), which corresponds to a temperature of 840 nK.

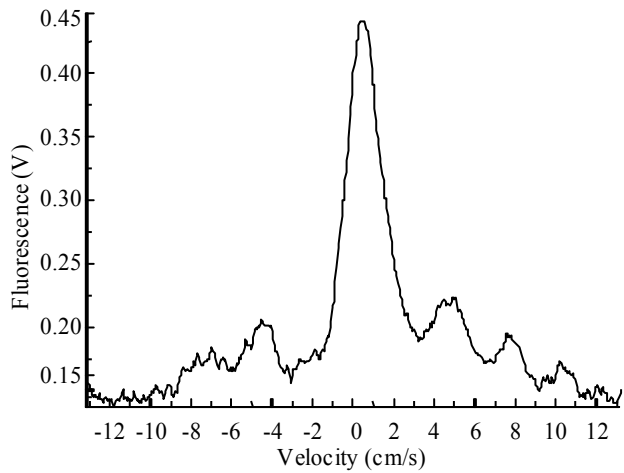

Figure 9: Velocity distribution after 5 ms second-stage cooling followed by 5 cycles of 15 μs pulses and 8 cycles of 25 μs pulses of red/green cooling. A fit of the narrow central feature to a Lorentzian yields a velocity (HWHM) of 1.0 cm/s (2/3 of a recoil) and a temperature of 520 nK.

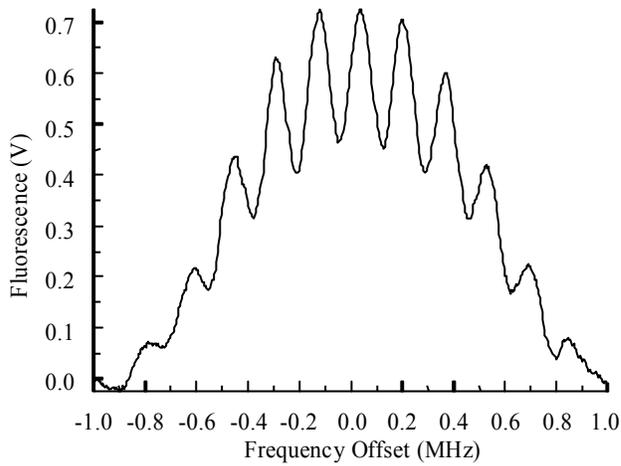

Figure 10: Two-pulse Ramsey fringes with 1 μs red pulses separated by a 5 μs Ramsey time. Fourier-transform-limited spectrum.